\documentclass[%
 reprint,%
 amssymb, amsmath,%
 aip,cha,%
]{revtex4-1}

\usepackage{bm}%
\usepackage{graphicx}
\usepackage[colorlinks=true,linkcolor=blue]{hyperref}%
\expandafter\ifx\csname package@font\endcsname\relax\else
 \expandafter\expandafter
 \expandafter\usepackage
 \expandafter\expandafter
 \expandafter{\csname package@font\endcsname}%
\fi
\usepackage{graphicx}
\usepackage{subfigure}
\usepackage{hyperref}
\usepackage{color}
\usepackage{epsfig}
\usepackage{verbatim}
\usepackage{epstopdf}

\begin{document}

\title{Emergent dynamics in delayed attractive-repulsively coupled networks}

\author{Prosenjit Kundu}
\affiliation{Department of Mathematics, National Institute of Technology, Durgapur 713209, India}
\author{Lekha Sharma}
\affiliation{Department of Mathematics, National Institute of Technology, Durgapur 713209, India}
\author{Mauparna Nandan}
\affiliation{Dr. B. C. Roy Engineering College, Durgapur 713206, India}
\author{Dibakar Ghosh}
\affiliation{Physics \& Applied Mathematics Unit, Indian Statistical Instiute, Kolkata 700108, India}
\author{Chittaranjan Hens}
\affiliation{Physics \& Applied Mathematics Unit, Indian Statistical Instiute, Kolkata 700108, India}
\author{Pinaki Pal}
\affiliation{Department of Mathematics, National Institute of Technology, Durgapur 713209, India}
 
%\author{AIP Journal Program}%
%\email{tex@aip.org}
%\affiliation{American Institute of Physics\\Suite 1NO1, 2 Huntington Quadrangle\\Melville, New York 11747-4502, USA}%

%\date{March 2010}%
%\revised{August 2010}%
\begin{abstract}
We investigate different emergent dynamics namely oscillation quenching and revival of oscillation in a  global network of identical oscillators coupled with diffusive (positive) delay coupling as it is perturbed by symmetry breaking localized repulsive delayed interaction. Starting from the oscillatory states (OS) we systematically identify three types of transition phenomena  in the parameter space: $(1)$ The system may reach inhomogeneous steady states (IHSS)  from the homogeneous steady state (HSS)  sometimes called as the transition from amplitude death (AD) to oscillation death (OD) state i.e. OS-AD-OD scenario,  $(2)$ Revival of  oscillation (OS)  from the AD state (OS-AD-OS) and $(3)$  Emergence of  OD state  from oscillatory state (OS)  without passing through AD i.e. OS-OD.  The dynamics of each node in the network is assumed to be governed either by identical limit cycle Stuart-Landau system or by chaotic R\"{o}ssler system. Based on clustering behavior observed in oscillatory  network we derive a reduced low-dimensional  model of the large network. Using the reduced  model, we investigate the effect of time delay on these transitions and demarcate  OS, AD and OD regimes in the parameter space.  We also explore and characterize the bifurcation transitions present in both systems.  The generic behavior of the low dimensional model and full network are found to match satisfactorily. 
\end{abstract}

\pacs{ 05.45.Xt, 05.45.Gg}

\maketitle
{\bf We systematically explore the combined impact of delay diffusive interaction and local repulsive links in globally coupled network of chaotic and limit cycle oscillators. For a certain delay, the system's oscillations are collapsed to steady states due to the presence of localized repulsive links. Depending on strength and links the oscillatory system may  directly move to  oscillatory state from amplitude death (AD) state i.e. homogeneous steady state (HSS) or may collapse into another set of fixed points (IHSS or OD) from AD. Based on a synchronization cluster behavior we squeeze the big network into two coupled models which can successfully mimic the full network. We have also identified  the bifurcation routes for revival of oscillations from steady states as well as the emergence of inhomogeneous steady states from homogeneous steady state.}

\section{Introduction}
%Oscillation quenching is one of the fundamental collective phenomena observed in complex systems consisting of large number of interacting units~\cite{G Saxena:Phys Rep_2012,A Koseska:Phys Rep_2013}. Although the phenomenon of oscillation quenching has long been known~\cite{rayleigh}, the topic  has received considerable attention of the researchers after it has been reported in coupled dynamical systems with parameter mismatch~\cite{PC Matthews:PRL_1990}. Later the oscillation quenching has been observed different conditions including time delay in coupling (see the reviews~\cite{G Saxena:Phys Rep_2012,A Koseska:Phys Rep_2013} and references therein). 

%It has been clearly demonstrated in~\cite{A Koseska:PRL_2013} that quenching of oscillation in coupled dynamical systems can manifest itself in two distinct forms namely amplitude death (AD) and oscillation death (OD).   

A combination of attractive and repulsive interaction  may create complex collective features  in a network of dynamical units ~\cite{Zanette2005,Strogatz-Hongprl}. For instance, the oscillatory networks may settle down to diverse steady states with finite  number of clusters~\cite{Chandrasekhar2018} or to  a combination of coherent and non-coherent states~\cite{Pikovsky,Mishra15, Chen2009}.  The synchronization of a brain network is also studied in a mixture of attractive-repulsive interaction~\cite{Rabinovich2006}. On the other hand, an attractively coupled oscillatory graph can be recast into a Turing type bifurcation~\cite{A Turing:PTRSL_1952} if a fraction of  nodes are controlled by localized repulsive feedback, a strategy successfully applied by breaking the symmetry of globally coupled limit cycles and chaotic oscillators \cite{Nandan2014}. The phenomena with same strategy is also explored for relay system \cite{Zhao2018}.   In a Turing type bifurcation, inhomogeneity is created from a homogeneous pattern for a certain range of diffusion. This is mapped as oscillation quenching mechanism~\cite{G Saxena:Phys Rep_2012, A Koseska:Phys Rep_2013} in oscillatory networks, where the system undergoes transition from homogeneous steady states (HSS) to inhomogeneous steady states (IHSS). Homogeneous steady state (HSS) is  referred as amplitude death(AD) and  IHSS is named as oscillation death (OD) state. 
The emergence of  AD occurs in a system of coupled oscillators when all the oscillators converge to a common fixed point or homogeneous steady state (HSS) \cite{Aronsen1990,Konishi2003,Karnatak2007,Matthews1990,Sen1998,Resmi2011} by resetting themselves into original equilibrium points.   On the other hand OD  state naturally depends on coupling strength and eventually the oscillators are splitted into  different steady states (IHSS)   \cite{A Koseska:Phys Rep_2013,A Koseska:PRL_2013,Hens2013}.
The  Turing like transition (OS-AD-OD) has been established in coupled oscillators with parameter mismatch~\cite{A Koseska:PRL_2013,Zou2013}, local repulsion \cite{Hens2013, Nandan2014}, mean field interaction \cite{Banerjee2014},  cyclic coupling \cite{Bera2016}, direct and indirect interactions \cite{pla2016} etc. and further validated with electronic experiments \cite{Banerjee2014_expt,Suresh2016} and recently Shrimali {et al.} discovered that the emergence of AD can be explosive \cite{Verma2017} in mean field diffusion interaction. On the other hand, it is already accepted that a delayed diffusion \cite{Sen1998}  or purely delayed  repulsive mean field \cite{Bera2016_pla} can systematically transform an oscillatory network to   AD and make it vice versa.
\par The key question, we raise here is  whether and under which condition a delayed attractive-repulsive interaction can break the symmetry of the system by  transforming   AD into OD, analogous to the occurrence of  instability in a purely homogeneous medium. More importantly, we study the emergence of symmetry breaking inhomogeneity by perturbing the  delayed diffusive network with delayed  repulsive local mean field.  We partially control or perturb (increasing the asymmetry up to an optimal level) the  network by systematically adding local repulsive links within the network  until  (AD-OD) transition is observed. Interestingly we have also shown that depending on the parameter space the network may be brought back to the oscillatory states (OS) from AD states avoiding OD states. In another parameter space, the system goes directly to OD state from  the OS state. Note that, the return or re-emergence of  oscillatory states from death are investigated by  setting a processing delay  in the coupling scheme \cite{Zou2013_prl} and  experimentally verified by introducing a feedback factor in diffusive system \cite{Zou2015, Nagao2016}.
  To explore these phenomena (OS-AD-OD, OS-AD-OS, OS-OD) we have used two paradigmatic nonlinear dynamical systems namely Stuart-Landau (SL) limit cycle oscillator and chaotic  R\"ossler system separately to model the dynamics of the individual nodes of the networks. For both cases we have considered a delayed global network of 
oscillators and gradually added local repulsive delayed links one by one and then perform numerical search when the above mentioned transitions appear in the network.  We have also shown that for a certain range of delay strength, the AD island can be expanded in the phase space of the coupling strength and the number of perturbed nodes (i.e. number of negative links). Further increase of  delay shrinks the AD or OD island, a consistent  feature appeared in both the oscillators.
Furthermore, based upon the two cluster synchronization which appears before the HSS, we are able to reduce the global  networks into two coupled systems, where a semi analytical treatment supports our numerical simulation. We have shown that the reduced system behaves quite similarly as its network counterpart.   

\section{Mathematical Model}
%\begin{eqnarray}
We consider a complete graph of size $N$, where the flow vector of each node is described by $f{\bf (X)}$. Assume that $p$ number of oscillators are affected by the localized negative links, in which the feed back comes from the unperturbed group of oscillators. We describe the perturbed population as   
%\vspace{-0.15cm}
\begin{eqnarray}
{\dot{\bf X}_{\it k}} &=& f({\bf X_{\it k}}(t)) + \frac{\epsilon}{N}\Gamma_1 \sum_{j=1}^{N}({\bf X_{\it j}}(t - \tau) - {\bf X_{\it k}}(t))\nonumber\\ 
              &-&\epsilon\Gamma_2({\bf X_{\it k}}(t) + {\bf X^*}(t - \tau)), { ({\it k}=1,2,...,{\it p})}
%\vspace{-0.15cm}
\label{eq:sub_population_1}
\end{eqnarray}
and the unperturbed population of the network is written as 
%\vspace{-0.15cm}
\begin{eqnarray}
\label{eq:sub_population_2}
{\dot{\bf X}_{\it l}} =f({\bf X_{\it l}}(t)) + \frac{\epsilon}{N}\Gamma_1\sum_{ j=1}^{N}({\bf X}_{j}(t - \tau) -{\bf X}_{l}(t)), \nonumber\\
(l=p+1,...,N), 
\end{eqnarray}
in which %$f :{\bf{R}}^m\rightarrow {\bf{R}}^m$ is a function and 
 $\Gamma_1$ and $\Gamma_2$ are $m\times m$ 
binary  matrices that encode the information of attractive and repulsive coupling variables.
% that represents all-to-all coupling. $\Gamma_2$ is another 
%$m\times m$ binary matrix that encodes  the information of the local repulsive interaction in which the number of nonzero 
%($1$) values is  $p$. 
The number of perturbed nodes is $p$ and that of the unperturbed nodes is $N -p$.
The symbol
 ${\bf X^*}$ 
 represents any arbitrary node of the  unaffected  population impacting the first population negatively but with some finite delay ($\tau$). The coupling component  
$\epsilon\Gamma_2 ({\bf X}_k + {\bf X^*})$ represents the additional repulsive link for a positive 
$\epsilon$. 
 Note that we have used a local repulsive mean field as a perturbative function, a common coupling scheme to create AD-OD transition in coupled oscillators \cite{Hens2013,Nandan2014}. If the system populates to amplitude death states then all the states are time independent fixed points i.e. ${\bf X_1=X_2=\dots X_p=X_{p+1}= \dots=X_N}$. 
On the other hand, in OD states the oscillators populate in nontrivial coupling  dependent fixed points where ${\bf X_1 \neq X_2 \neq \dots X_p \neq X_{p+1} \neq \dots \neq X_N}$ or in the case of two cluster steady states  ${\bf X_1 = X_2 = \dots = X_p \neq X_{p+1} = \dots = X_N}$.    The choices of  
$\Gamma_1$ and $\Gamma_2$ 
are wide and can be chosen in diverse ways. For instance, for Stuart-Landau oscillators ($m=2$) we will use use the attractive delay coupling in $x$  variable i.e. 
$\Gamma_1=\begin{pmatrix}
  1 & 0 \\ 
  0 & 0
\end{pmatrix} 
$
and  
$\Gamma_2$ is chosen a such a way that repulsive coupling is used in $y$ variable,
therefore $\Gamma_2=\begin{pmatrix}
  0 & 0 \\ 
  0 & 1
\end{pmatrix} 
$. 
The coupling scheme is  robust and generic such that reversing the coupling variables can create amplitude or oscillation death states from the steady oscillations.   
Keeping similar configuration as described here, in the subsequent subsections, we investigate the phenomenon of transition from AD to OD in networks of limit cycle as well as chaotic oscillators. 
\vspace{-0.15cm}
\section{Transition in network of {Stuart-Landau} oscillators}
We start with an all-to-all network of paradigmatic Stuart-Landau (SL) limit cycle oscillators. The complex version of SL oscillator can be written as  $\dot{z}=[1+iw-{{z}}^2]z$, where $w$ is the intrinsic frequency and the complex variable $z = x + iy$ with real variables $x$ and $y$. Following the coupled equations described in  (\ref{eq:sub_population_1} - \ref{eq:sub_population_2})
we can write the two groups of SL oscillators (by decomposing into real and imaginary part) as,
\ \vspace{-0.05cm}
\begin{eqnarray}
\label{eq:reduced_population_1}
{\dot{x}_k}&=&[1 - (x_k(t)^2 + y_k(t)^2)]x_k - \omega y_k(t)\nonumber\\
&+&\frac{\epsilon}{N}\sum_{j=1}^{N}(x_j(t - \tau) - x_k(t)), \\
{\dot{y}_k}&=&[1 - (x_k(t)^2 + y_k(t)^2)]y_k + \omega x_k(t) \nonumber\\
&-&\epsilon(y_k(t) + y_N(t - \tau)), (k = 1, 2, \dots, p),\nonumber
\vspace{-.5cm}
\end{eqnarray} 
and
\begin{eqnarray}
\label{eq:reduced_population_2}
{\dot{x}_{\it l}}&=&[1 - (x_{\it l}(t)^2 + y_{\it l}(t)^2)]x_{\it l}(t) - \omega y_{\it l}(t)\nonumber\\
&+&\frac{\epsilon}{N}\sum_{j=1}^{N}(x_j(t -\tau) - x_{\it l}(t)),\\
{\dot{y}_{\it l}}&=&[1 - (x_{\it l}(t)^2 + y_{\it l}(t)^2)]y_{\it l}(t) + \omega x_{\it l}(t), \nonumber \\
&&(l = p+1, p+2, \dots, N). \nonumber
\end{eqnarray}
\subsection{Numerical Simulation}
We first perform numerical simulation of the equations (\ref{eq:reduced_population_1})-(\ref{eq:reduced_population_2}) using DDE23 solver of MATLAB for $N = 50$. We then perturb $20$ nodes of the network with local repulsive interaction i.e. $p = 20$. We set the delay strength at $0.5$ for both diffusive and repulsive interaction. 
\begin{figure}[h]
%\centerline{\epsfxsize=4.0in\epsfbox{./chapter4/r1_r2_100_200_300_400.eps}}
\includegraphics[height=7cm,width=9cm]{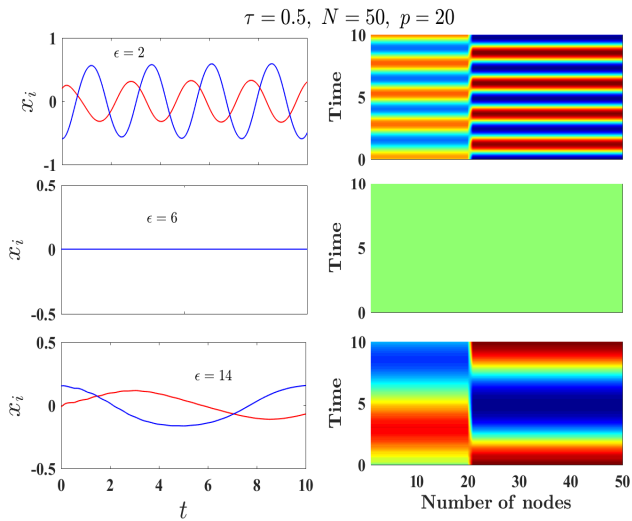}
\caption{Temporal and spatio-temporal evolution of the variable $x_i (i = 1,2, \dots, 50)$ for an all-to-all coupled network of $50$ SL oscillators with a delay $\tau=0.5$ and 
$p=20$ as obtained from numerical simulation. Left column describes the time evaluation of the variables $x_i (i = 1,2, \dots, 50)$ for three values of $\epsilon$. Solid red and blue curves represent the time evolution of the variable corresponding to perturbed and unperturbed nodes respectively. In the right column,  corresponding spatio-temporal pattern of all nodes are shown.} 
\label{fig:LC_AD_OD_fix_tau_1}
\end{figure}
The results are shown in the 
Fig.\ \ref{fig:LC_AD_OD_fix_tau_1}. We consider three coupling strengths $\epsilon=2,6,14$. 
The left column represents the temporal evolution of all the nodes and the spatio-temporal propagation of all the nodes are displayed in the right column. In the lower coupling range ($\epsilon=2$), the whole population splits into two synchronized oscillatory states (OS) as shown in the upper panel. We  observe that the time evolution of the  perturbed population (first $20$ nodes) has less amplitude (shown with red color in the left column) than the unperturbed nodes (shown in blue color). The spatio-temporal propagation of the whole population is shown in the right column of the upper panel where two clusters are clearly observed. If we increase the coupling strength ($\epsilon=6$), the whole population collapses into homogeneous steady states (HSS),  a trivial zero  equilibrium as shown in the middle panel. If we now almost double the coupling strength ($\epsilon=14$), 
the oscillatory nature (OS state) revives as logical consequence of high competition between attractive and repulsive interaction. Keeping the number of repulsive links constant, the system undergoes a transition from oscillatory (OS) to HSS/AD state and then again returns to oscillatory state (OS) for higher coupling strength. 
\begin{figure}[h]
\includegraphics[height=7cm,width=9cm]{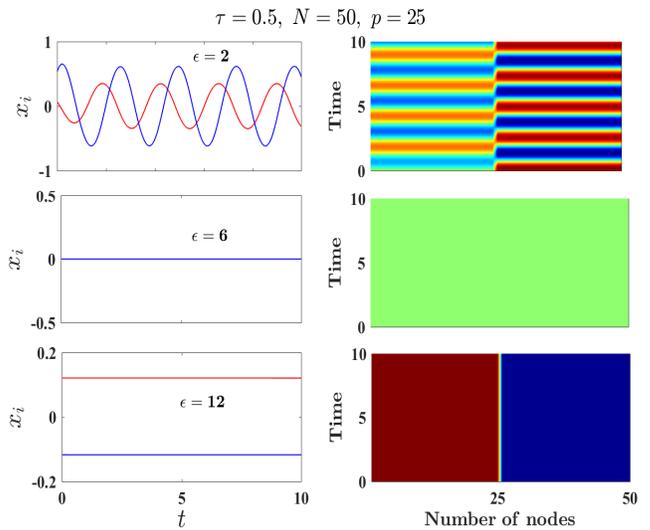}
\caption{Time series (left panel) and corresponding spatio-temporal patterns (right panel) of the variables $x_i (i = 1,2, \dots, 50)$ as obtained from numerical simulation of a network of $50$ SL oscillators where $50\%$ of oscillators are perturbed ($p = 25$) for three values of the coupling strength $\epsilon$. Solid red and blue curves in the left panel corresponds to the time series of the perturbed and unperturbed nodes of the network respectively.} 
\label{simu2}
\end{figure}

Next we perform numerical simulation of the network of size $N = 50$ coupled SL oscillators having $p = 25$ i.e. $50\%$ nodes of the network are perturbed with repulsive links. The temporal evolution of the 
$x$ 
variables as well as the spatio-temporal patterns of the observed dynamics is shown in the Fig.\ \ref{simu2}. From the figure we observe that as the coupling strength 
$\epsilon$ 
is increased, the network state move from oscillatory to AD state and for further increase of 
$\epsilon$, 
a transition from AD to OD occurs. Note that oscillatory state is not revived at higher coupling strength in this case. 
\begin{figure}[h]
\includegraphics[height=7cm,width=9cm]{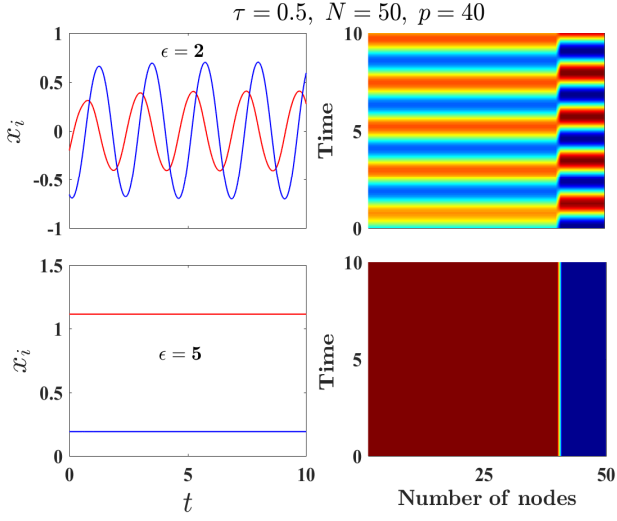}
\caption{Time series (left panel) and corresponding spatio-temporal patterns (right panel) as obtained from numerical simulation of a network of $50$ SL oscillators where $80\%$ of oscillators are perturbed ($p = 40$) for two values of the coupling strength $\epsilon$. Solid red and blue curves in the left panel corresponds to the time series of the perturbed and unperturbed nodes of the network respectively.} 
\label{simu3}
\end{figure}
A comparatively large perturbation ($80\%$ repulsive links) transforms the oscillatory population (OS) into OD state, a direct transition (Fig.\ \ref{simu3}) appears resulting an absence of HSS/AD state. Depending on the network perturbation (percentage of negative links) we can reach three possible transition scenario: OS-AD-OS, OS-AD-OD or OS-OD. 

From the numerical simulation,  it is clear now  that   for  a given delay ($\tau = 0.5$), as the percentage of repulsive links is increased, the system may undergo various types of transition (OS-AD-OS, OS-AD) and 
the splitting of the populations into two oscillatory clusters (before the  onset of steady states)  enables us to derive a low dimensional description of the network. The details are described in the next subsection.
\noindent

\begin{figure}[h]
\includegraphics[height=6cm,width=7cm]{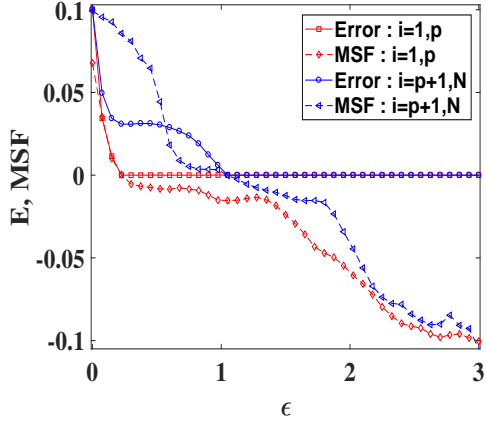}
\caption{Synchronization Error (E) and master stability function (MSF) of two populations as a function of coupling strength $\epsilon$ for the network of limit cycle oscillators.} 
\label{MSF_Error_LS}
\end{figure} 
\subsection{A reduced order model}
From the time evolution of the  $x$-variable for
 $\epsilon = 2$ 
 as shown in Figs.~\ref{fig:LC_AD_OD_fix_tau_1},~\ref{simu2} and~\ref{simu3}, we observe that the nodes with repulsive link and the remaining nodes without repulsive links form two distinct clusters and each cluster is in perfect synchrony.  Due to the presence of 2-cluster synchronization we may write $x_1=x_2=...= x_p=X_1$ for ($i = 1,2, \dots, p)$   and  
$x_{p+1} = x_{p+2} =...= x_N =X_2$. The same can be applied to the $y$ variable as the clustering behavior is independent of variables.  Inserting  these cluster relations in Eqs.~(\ref{eq:reduced_population_1}) and~(\ref{eq:reduced_population_2}),   we arrive at the following low dimensional representation \cite{Daido2004,Nandan2014,Zou2009,Kundu2018} of the diffusively coupled (but locally perturbed by delayed repulsive link) network:
\begin{eqnarray}
\label{eq:reduced_popu_final1}
{\dot{X}_1}&=&[1 - (X_1(t)^2 + Y_1(t)^2)]X_1(t) - \omega Y_1(t)\nonumber\\
&+&  \frac{q\epsilon}{N}(X_2(t - \tau)-X_1(t)) \nonumber\\
&+& \frac{p\epsilon}{N}(X_1(t - \tau)-X_1(t)), \nonumber\\
\dot{Y_1}&=&[1 - (X_1(t)^2 + Y_1(t)^2)]Y_1(t) + \omega X_1(t) \nonumber\\
&-& \epsilon(Y_1(t) + Y_2(t - \tau)), \\
{\dot{X}_2}&=&[1 - (X_2(t)^2 + Y_2(t)^2)]X_2(t) - \omega Y_2(t)\nonumber\\
&+&  \frac{p\epsilon}{N}(X_1(t - \tau)-X_2(t)) \nonumber\\
&+& \frac{q\epsilon}{N}(X_2(t - \tau)-X_2(t)), \nonumber\\
\dot{Y_2}&=&[1 - (X_2(t)^2 + Y_2(t)^2)]Y_2(t) +\omega X_2(t).\nonumber
\end{eqnarray}
\begin{figure}[h]
\includegraphics[height=7cm,width=9cm]{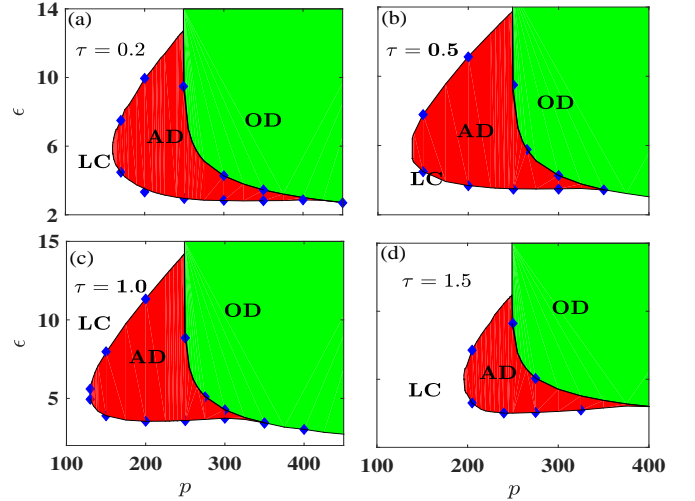}
\caption{Phase diagram in $\epsilon - p$ plane as obtained from the reduced model of the network of SL units ($N=500$, $\omega = 3.0$) for four values of $\tau$. Limit cycle, AD and OD regions  are denoted by white, red and green colors respectively. Different boundaries separating these regions obtained from the numerical simulation of the full network are shown with blue diamonds.} 
\label{fig:LC_AD_OD_fix_tau}
\end{figure}
  A numerical observation of two cluster synchronization demands stability under certain linear perturbation. This is described in details in appendix for a generalized vector models (\ref{eq:sub_population_1}) and (\ref{eq:sub_population_2}).   Note that, based on these coupled equations and reduced coupled equations (Eqns.\ (\ref{msf1}),(\ref{msf2}), (\ref{App_red1}),  (\ref{App_red2}) in appendix), we calculate the maximum Lyapunov exponent (master stability functions i.e. MSF) of the transverse manifold of each cluster. Sign changes (From positive to negative)  in MSF ensures the stability of each cluster. It is clear from the Fig.\ \ref{MSF_Error_LS} that the perturbed cluster becomes synchronized earlier (Shown by red color), however unperturbed population takes more strength (shown in blue) to synchronize. Both cluster synchronize at 
$\epsilon \sim 1 $. A time average mean Eucledian distances among the nodes (within  clusters) are also plotted as Error (E) function which closely fit with MSF curves.
We have calculated MSF for the network size  $50$ and  repulsive links  $25$ and it is consistent with the first row of the Fig.\ \ref{simu2} where two cluster states appear at $\epsilon =2.0$.

Based on these equations, we explore the parameter space 
$\epsilon-p$ 
for four different delay ($\tau$) values. Figure~\ref{fig:LC_AD_OD_fix_tau} shows the phase diagram constructed from the low dimensional model for a large network 
($N = 500$) 
where different regions (AD, OD and limit cycle) are clearly demarcated for four values of 
$\tau$. For 
$\tau=0.2$ (Fig.~\ref{fig:LC_AD_OD_fix_tau}(a)),  we observe that in weak coupling ($\epsilon \sim 0$) 
and less repulsive environment 
($p \sim 100$) 
the network exhibits oscillatory e.g limit cycle behavior (LC) shown in white space. A moderate increase of 
$p$ and $\epsilon$ 
sets the whole population into  AD state as shown by red color. Further increase of $p$, the system reaches to inhomogeneous steady states OD as shown in green color. The same type of transition is observed  for higher delayed interaction 
($\tau=0.5$ and $\tau=1.0$) 
as shown in (Fig.\ \ref{fig:LC_AD_OD_fix_tau}(b)-(c)).  We have also observed that AD island is slightly enhanced for the increase of delay upto 
$1$. For higher delay ($\tau=1.5$) the AD island is decreased significantly. Higher delay controls the regime of  oscillatory state by increasing its area in $\epsilon-p$ space. Although
OS-AD-OD, OS-AD-OS and  OS-OD are still visible in vertical as well as in horizontal direction  shown in (Fig.\ \ref{fig:LC_AD_OD_fix_tau}(d)).  OS-AD-OS transition occurs at lower 
$p$ 
(around $p\sim200$) 
where as OS-OD transition occurs when the $80\%$ 
nodes are inherently perturbed by the delayed repulsive links. \\
Next we perform numerical simulation in the whole  network  to validate the results obtained from the  low dimensional model. We  use two  global order parameters defined in \cite{Nandan2014} to distinguish the AD or OD states from its oscillatory counter part. Considering few values from 
$\epsilon - p$ 
phase space we numerically identify whether the whole system is  either in AD or OD states. The boundaries in each cases  determined from the large network have been shown with blue diamonds in Fig.~\ref{fig:LC_AD_OD_fix_tau}. It is clearly observed that the numerical  results are perfectly matched with the results obtained from the low dimensional model. 

\begin{figure}[h]
\includegraphics[height=!,width=9cm]{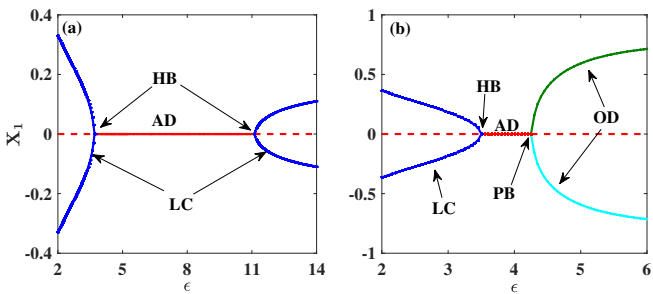}
\caption{Bifurcation diagrams constructed from the reduced order model~(\ref{eq:reduced_popu_final1}) of a network of size $N = 500$ with delay $\tau = 0.5$ for (a) $p = 200$ and (b) $p=300$. Along horizontal and vertical axes the variation of coupling strength $\epsilon$ and extremum values of a system variable $X_1$ have been shown respectively. Solid blue and red curves represent stable limit cycles (LC) and AD states respectively, while dashed red curves represents  unstable zero fixed points. OD states are represented with solid cyan and green curves.} 
\label{bifurs_ls}
\end{figure}

  Next we analytically calculate the critical curves for AD regions from the reduced model (\ref{eq:reduced_popu_final1}). For linear stability of AD state, we have $X_1=X_2=Y_1=Y_2=0$ is the fixed point of the reduced model (\ref{eq:reduced_popu_final1}).  The characteristic equation of Eq.~(\ref{eq:reduced_popu_final1}) at the trivial fixed point is given by $det(L)=0$, where
	%\begin{widetext}
	\begin{eqnarray}
	L=\begin{bmatrix}
	1-\frac{q\epsilon}{N} -\lambda      & -w & \frac{q\epsilon}{N}e^{-\lambda \tau} & 0 \\
	+ \frac{p\epsilon}{N}(e^{-\lambda \tau}-1) &~ &~ & \\
	\\
	
	w       & 1-\epsilon-\lambda & 0 & -\epsilon e^{-\lambda \tau} \\
	\\
	\frac{p\epsilon}{N}e^{-\lambda \tau} & 0 & 1-\frac{p\epsilon}{N}-\lambda & -w\\
	~& ~& + \frac{q\epsilon}{N}(e^{-\lambda \tau}-1) & \\
	\\
	
	0       & 0 & w &  1-\lambda 
	\end{bmatrix}
	\nonumber
	\end{eqnarray}
	%\end{widetex}
	and $\lambda$ is the characteristic root or eigenvalue. Eventually $det(L)=0$ takes the form 
	\begin{eqnarray}
	P(\lambda) + Q(\lambda)e^{-\lambda \tau} +R(\lambda)e^{-2\lambda\tau} =0,
	\label{Ch_eq}
	\end{eqnarray}
	where,
	\begin{eqnarray}
	P(\lambda) &=& -\lambda^4+(4-3\epsilon)\lambda^3 +(w-2w^2-2+2\epsilon-\epsilon^2)\lambda^2 \nonumber \\
	&+&(4w^2-3w^2\epsilon)\lambda -(w^4+w^2(1-\epsilon)(2-\epsilon)+(1-\epsilon)^3), \nonumber \\
	Q(\lambda) &=&\epsilon\lambda^3-(\lambda^2-\epsilon+1)\lambda^2+w^2\epsilon\lambda+\frac{w^2p\epsilon^2}{N} -w^2\epsilon\nonumber \\
	&-&\epsilon(1-\epsilon),\\
	R(\lambda)&=&\frac{\epsilon^2p}{N}[w^2+\frac{q}{N}(\epsilon-2)]\lambda\nonumber . 
	\end{eqnarray}
	Now we set the real part of the eigenvalue equals to zero and consider $\lambda_2$ be the imaginary part of the eigenvalue i.e. $\lambda=i \lambda_2$ in Eq.\ (\ref{Ch_eq}).
	Separating real and imaginary parts, we get
	\begin{eqnarray}
	(a+\epsilon(1-\epsilon)\lambda_2^2)\cos(\lambda_2\tau) +(w^2\epsilon\lambda_2-\epsilon\lambda_2^3)\sin(\lambda_2\tau) \nonumber \\
	+b\cos(2\lambda_2\tau)+\lambda c\sin(2\lambda_2\tau)\nonumber \\
	=\lambda_2^4 +(w-2w^2+2\epsilon-\epsilon^2 -2)\lambda_2^2 \nonumber \\
	+ (w^4+w^2(1-\epsilon)(2-\epsilon)+(1-\epsilon)^3),
	\label{real_part}
	\end{eqnarray}
	and \begin{eqnarray}
	(a+\epsilon(1-\epsilon)\lambda_2^2)\sin(\lambda_2\tau) +(w^2\epsilon\lambda_2-\epsilon\lambda_2^3)\cos(\lambda_2\tau)\nonumber \\
	+b\sin(2\lambda_2\tau)+ \lambda c\cos(2\lambda_2\tau)
	=-\lambda_2^3(4-3\epsilon)\nonumber \\+(4w^2-3w^2\epsilon)\lambda_2,
	\label{im_part}
	\end{eqnarray}
	where $a=\frac{w^2p\epsilon^2}{N}-w^2\epsilon-\epsilon(1-\epsilon)$, $b=\frac{w^2p\epsilon^2}{N}$ and $c=\frac{\epsilon^2pq}{N^2}(\epsilon-2)$.
	The critical  AD curves in the $p-\epsilon$ parameter plane obtained from the set of equations (\ref{real_part}) and (\ref{im_part}) for different values of the time-delay $\tau$ are shown in Fig.\ \ref{fig:LC_AD_OD_fix_tau} by the black lines. The analytically derived linear stability curves for AD state are in excellent agreement with the numerical results from reduced model as well as large network. 
To understand the route to the oscillation quenching mechanism, we construct two qualitatively different bifurcation diagrams from the low dimensional model with the help of XPPAUT software~\cite{auto}. The first type of transition (OS-AD-OS) is confirmed   in Fig.~\ref{bifurs_ls} (a), in which  the model parameters are taken as $N = 500$, $p = 200$ and $\tau = 0.5$.   
AD appears from the oscillatory system via reverse Hopf bifurcation (HB) at 
$\epsilon = 3.7$ and it  exists in the range $3.7 \leq \epsilon \leq 11.1$ further increase of 
$\epsilon$ 
the oscillation reappears through forward Hopf bifurcation. 

For a higher values of $p$ ($p=300$), the network shows a transition from oscillatory to AD and then from AD to OD (OS-AD-OD)   shown in Fig.~\ref{bifurs_ls}(b). From the bifurcation diagram, we observe that the periodic oscillatory solutions (solid blue curves) exists in the range $2 \leq \epsilon \leq 3.5$ and at $\epsilon = 3.5$ the zero fixed point becomes stable via reverse Hopf bifurcation and AD state is originated in the system (solid red curve in the Fig.~\ref{bifurs_ls}(b)). The AD state is stable in the range $3.5 \leq \epsilon \leq 4.25$. At $\epsilon = 4.25$, the AD state becomes unstable via supercritical pitchfork bifurcation (PB) and OD states appears. Two stable stationary branches appear from $\epsilon = 4.25$ which are shown with solid green and cyan curves in the bifurcation diagram.  Note that, the these two bifurcation diagrams are the  description of a reduced model in a specific parameter space which is completely consistent with the transition occurs in large network.
 
\section{Network of chaotic oscillators}
We extend our observation on AD-OD transition to a network of chaotic R\"ossler oscillators. We separate all the globally coupled R\"ossler oscillators once again into two sub-populations, one perturbed by repulsive links and another unperturbed. The perturbed and unperturbed groups of R\"ossler oscillators are given by 

\begin{eqnarray}
\label{eq:ross_popu_2}
{\dot{x}_k}&=&- y_{\it k}- z_{\it k} + \frac{\epsilon}{N}\sum_{j=1}^{N}(x_j(t-\tau) - x_{\it k}(t)),\\
{\dot{y}_k}&=&x_{\it k} + ay_{\it k}-\epsilon(y_k(t-\tau) + y_N(t)),\nonumber\\
{\dot{z}_{\it k}}&=&bx_{\it k} + z_{\it k}(x_{\it k} -c), \nonumber
\vspace{-0.25cm}
\end{eqnarray}
and
\begin{eqnarray}
\label{eq:ross_popu_1}
{\dot{x}_{\it l}}&=& - y_{\it l}- z_{\it l} + \frac{\epsilon}{N}\sum_{j=1}^{N}(x_j(t-\tau) - x_{\it l}(t)),\\
{\dot{y}_{\it l}}&=&x_{\it l} + ay_{\it l},\nonumber\\
{\dot{z}_{\it l}}&=&bx_{\it l} + z_{\it l}(x_{\it l} -c), \nonumber
\end{eqnarray}
respectively where $a$, $b$ and $c$ are a system parameters, $k=1, 2,...,p$ and $l=p+1, p+2,...,N$.
% \begin{figure}[h]
%\includegraphics[height=!,width=9cm]{rossler_attractive_repulsive_eps_tau_error.eps}
%\caption{Error of two sub population from the synchronized state plotted with the coupling strength for the network of chaotic  R\"ossler oscillators. } 
%\end{figure}
% \begin{figure}[h]
%\includegraphics[height=!,width=9cm]{rossler_attractive_repulsive_error_final_new.eps}
%\caption{} 
%\end{figure}
\subsection{Numerical Simulation}
\begin{figure}[h]
\includegraphics[height=7cm,width=9cm]{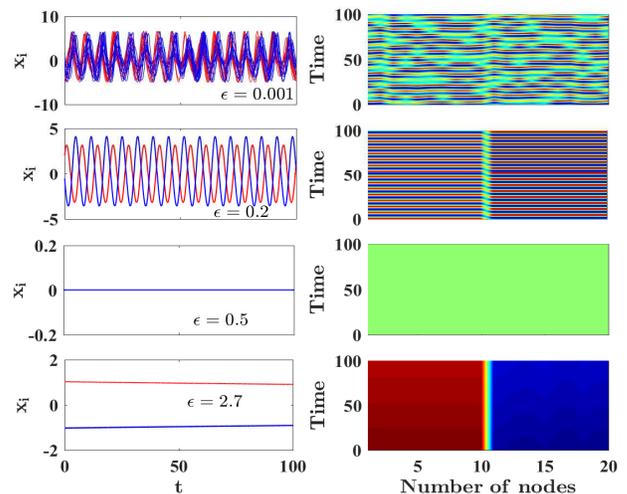}
\caption{Time series (left panel) and spatio-temporal (right panel) dynamics of the variables $x_i (i = 1,2, \dots, 20)$ for a global network of $20$ chaotic R\"ossler oscillators with $\tau = 0.5$ and $p = 10$. Solid red and blue curves in the left panel correspond to the perturbed and unperturbed nodes.} 
\label{simu_c}
\end{figure}

Similar to network of limit cycle oscillators, here also we perform numerical simulation of a small network of size $N = 20$ and number of negative links $p = 10$ for a set of values of the coupling constant $\epsilon$. The values of the parameters $a$, $b$ and $c$ are taken to be $0.36$, $0.4$ and $4.5$ respectively for which the system show chaotic behavior. Figure~\ref{simu_c} shows the results of the simulation. From the first column of Fig.\ \ref{simu_c}, we find that as the coupling strength increases the dynamics of the network passes through chaotic, periodic, amplitude death and oscillation death state respectively.  As expected, for networks of delay coupled chaotic oscillators perturbed by $50\%$ repulsive links, a transition from AD to OD state occurs (OS-AD-OD). We now explore this transition phenomena in detail by constructing a reduced model of the network. 

\subsection{A reduced order model}
\begin{figure}[h]
\includegraphics[height=!,width=6cm]{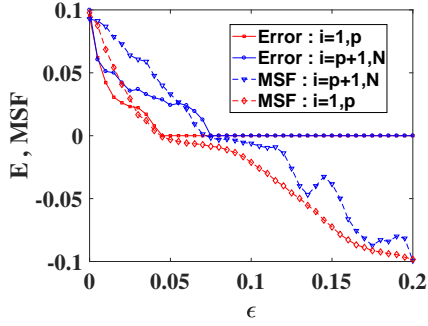}
\caption{Synchronization Error (E) and master stability function (MSF) of two populations as a function of coupling strength $\epsilon$ for the networks (\ref{eq:ross_popu_2}) and (\ref{eq:ross_popu_1}) of chaotic oscillators.} 
\label{MSF_Error_Chaos} 
\end{figure}

We adopt similar approach to derive a reduced order model of the network of chaotic R\"ossler oscillators. The formation of two groups is evident from the second row of Fig.~\ref{simu_c} before the onset of AD in the system, as it has occurred in the case of the network of limit cycles SL oscillators.  We can also check the stability of two cluster states by calculating the master stability functions from Eqns.\ (\ref{msf1}) and (\ref{msf2}) (Please see the Appendix).  As expected,  both clusters  are stable  beyond the coupling strength $\epsilon \sim 0.08$ (Fig.\ \ref{MSF_Error_Chaos})  which also ensures the stability of the numerically simulated two clusters shown in the second row of the Fig.\ \ref{simu_c} for $\epsilon=0.2$. 
 Therefore, by setting $(x_i, y_i, z_i) = (X_1,Y_1, Z_1), (i = 1,2, \dots, p)$ and  $(x_i, y_i, z_i) = (X_2, Y_2, Z_2), (i = p+1,p+2, \dots, N)$ and using the equations~(\ref{eq:ross_popu_2}) and~(\ref{eq:ross_popu_1}) we arrive at the following set of six equations

\begin{eqnarray}
\label{eq:reduced_popu_final}
{\dot{X}_1}&=&-Y_1 - Z_1 +   \frac{q\epsilon}{N}(X_2(t - \tau)-X_1(t)) \nonumber\\
&+& \frac{p\epsilon}{N}(X_1(t - \tau)-X_1(t)), \nonumber \\
{\dot{Y}_1}&=&X_1+aY_1 - \epsilon(Y_1(t-\tau) + Y_2(t)), \nonumber\\
{\dot{Z}_1}&=&bX_1+Z_1(X_1-c), \nonumber\\
{\dot{X}_2}&=&-Y_2 - Z_2 +  \frac{p\epsilon}{N}(X_1(t - \tau)-X_2(t)) \nonumber\\
&+& \frac{q\epsilon}{N}(X_2(t - \tau)-X_2(t)), \nonumber \\
{\dot{Y}_2}&=& X_2+aY_2, \nonumber\\
{\dot{Z}_2}&=& bX_2+Z_2(X_2-c),
\end{eqnarray}
which is expected to represent the dynamics of the entire network for the investigation of the transition  either OS-AD-OD, OS-AD-OS or OS-OD.
\begin{figure}[h]
\includegraphics[height=!,width=9cm]{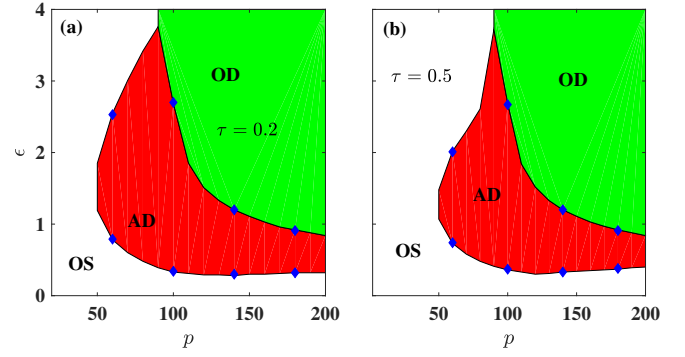}
\caption{Phase diagram in $\epsilon-p$ plane constructed from the reduced model of the network of chaotic R\"ossler oscillators of size $N=200$ ($a = 0.36$, $b =0.4$, and $c = 4.5$) for two values of $\tau$. White region represents oscillatory (OS) regime (either chaotic or periodic), AD and OD regimes are shown with red and green colors respectively. The blue diamonds shows the boundary points obtained from the numerical simulation of the full network.} 
\label{fig:LC_AD_OD_Rossler}
\end{figure}
We now use this model to investigate the transition phenomenon in the network of chaotic oscillators perturbed by repulsive links one by one. We consider a large network of R\"ossler oscillators of size $N = 200$ and $c = 4.5$. Figure~\ref{fig:LC_AD_OD_Rossler} shows the phase diagram in $\epsilon - p$ plane for two different values of $\tau$ as constructed from the low dimensional model~(\ref{eq:reduced_popu_final}). In the figure, oscillatory (OS), AD and OD regions are clearly demarcated with different colors. Note that OS regime consists of chaotic as well as periodic states of the system. However, chaotic regimes are very thin and observed only for very small values of 
$\epsilon$. 
 It is interesting to note here that we do not observe any direct transition from OS to OD states in the network of chaotic R\"ossler oscillators as $\epsilon$ 
 is increased for fixed value of 
 $p$ but one can observe a transition from OS to OD state as 
 $p$ is increased for high value of $\epsilon$. 
 
\begin{figure}[h]
\includegraphics[height=!,width=9cm]{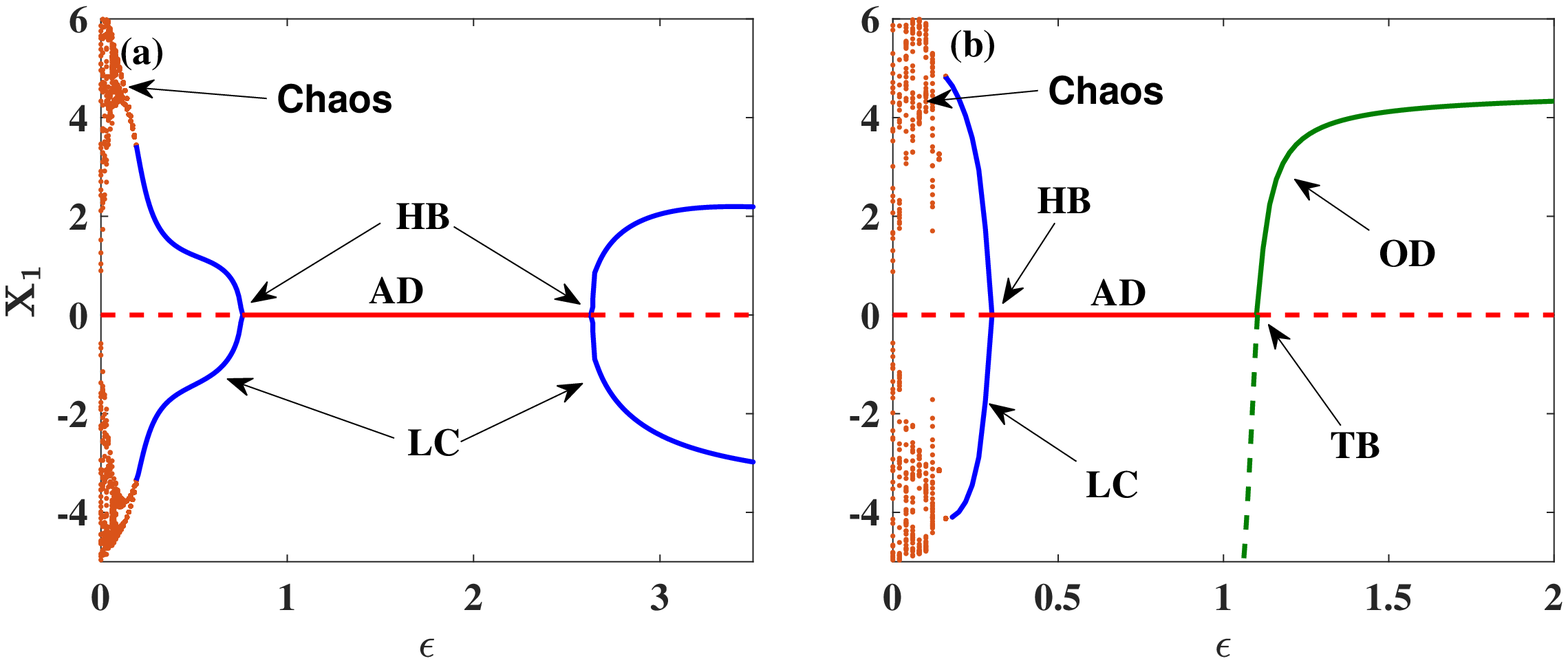}
\caption{Bifurcation diagrams constructed from the reduced order model~(\ref{eq:reduced_popu_final}) of a network of size $N = 200$ with delay $\tau = 0.2$ for (a) $p = 60$ and (b) $p=150$. Along horizontal and vertical axes the variation of coupling strength $\epsilon$ and extreme values of a system variable $X_1$ have been shown respectively. Solid blue and red curves represent stable limit cycles (LC) and AD states respectively, while dashed red curves represents  unstable zero fixed points. Stable OD states are represented with solid green curves and it is originated via transcritical bifurcation (TB). Brown dots represent chaotic solutions.} 
\label{bifurs_ro}
\end{figure}

To understand the transition routes observed in the network, we perform the bifurcation analysis of the low-dimensional  model~(\ref{eq:reduced_popu_final}). In order to understand the two qualitatively different transitions in the network we have perturbed the network with slightly less number of repulsive links 
($p=60$) and with moderately high number of negative links 
($p=150$).  
Figure~\ref{bifurs_ro} shows two bifurcation diagrams computed from the reduced order model~(\ref{eq:reduced_popu_final}) for $N = 200$, $\tau = 0.2$ and for two different values of $p$. The bifurcation diagram~\ref{bifurs_ro}(a) shows a transition from oscillatory state (Chaotic and periodic) to AD state for smaller number of negative links ($p = 60$). The AD state appears in the system via reverse Hopf bifurcation at $\epsilon = 0.76$. The AD state is stable in the range $0.76\leq \epsilon\leq 2.63$ and it becomes unstable via forward Hopf bifurcation (HB) at $\epsilon = 2.63$. After which the oscillation is again revived in the system. 
For a larger number of negative links ($p = 150$), a transition from AD to OD occurs. This is nicely depicted in the bifurcation diagram~\ref{bifurs_ro}(b). The AD state (solid red) is found to appear via reverse Hopf bifurcation at $\epsilon = 0.3$ and remain stable in the range $0.3\leq \epsilon \leq 1.1$. After that, OD states appear via transcritical bifurcation (green curves in the Fig.\ \ref{bifurs_ro}(b)) which is similar to the one as observed in absence of delay~\cite{Nandan2014}.

\section{Conclusions}
In this paper, we have investigated the effect of time delay on transition from OS-AD-OD, OS-AD-OS or OS-OD in globally coupled networks of nonlinear dynamical systems. The dynamics of each node of the network is assumed to be governed either by identical limit cycle Stuart-Landau system or by chaotic R\"{o}ssler system which interact with other nodes via delay diffusive coupling while some oscillators are perturbed through delay repulsive coupling. Numerical simulations of the network reveal that for a given number of repulsive links the whole population of the network forms two groups as the coupling strength crosses a threshold. The nodes with repulsive links form one group and the remaining nodes form another group. Each group of nodes is found to be in synchronized state  which is further confirmed by constructing the master stability functions. For further increase in the coupling strength, when the number of repulsive links is greater than a critical value, oscillation quenching either in the form of AD or OD appears in the system. From this information, we derive reduced order models of the large network. Using the reduced order model, we investigate the effect of time delay on the transition from AD to OD in the network and then demarcate the oscillatory, AD and OD regimes in the parameter space quite easily in detail. Finally we compare the reduced order model results with that of the direct numerical simulations performed with larger networks. The results of the reduced order model and the entire network are found to match satisfactorily. Moreover, we have performed bifurcation analysis of the reduced order models of the networks which proves to be very effective in finding the routes to oscillation quenching. 

\begin{acknowledgments}
Authors thank Manojit Ghosh  for the fruitful discussions. The authors also like to thank the anonymous reviewers for their fruitful suggestions. PK acknowledges support from  DST, India under the DST-INSPIRE scheme (Code: IF140880). CH is supported by INSPIRE-Faculty grant:  DST/Inspire/04/2017/003252. 
%(Code: IFA17-PH193). 
\end{acknowledgments}

\section{Appendix: Stability of synchronized clusters}
We analytically derive the condition for synchronization of Eqs.\ (\ref{eq:sub_population_1}) and (\ref{eq:sub_population_2}) by master stability function approach \cite{msf}. If  ${\bf X_f}$ and ${\bf X_s}$ are the synchronization manifolds for the two sub-populations, then ${\bf X_{\it k}=X_f}$ for $k=1, 2,...,p$ and ${\bf X_{\it l}=X_s}$ for $l=p+1, p+2, ...,N$. We Consider small perturbations ${\bf \delta X_{\it k}}$ and ${\bf \delta X_{\it l}}$ near the synchronization manifolds, i.e. ${\bf X_{\it k}=X_f+\delta X_{\it k}}$ for $k=1,...,p$ and ${\bf X_{\it l}=X_s+\delta X_{\it l}}$ for $l=p+1,...,N$. Then the variational equations corresponding to Eqs.(\ref{eq:sub_population_1}) and (\ref{eq:sub_population_2}) are respectively,
\begin{eqnarray}
\delta {\dot{\bf X}_{\it k}}&=&{\bf J}f({\bf X_{\it f}})\delta {\bf X_{\it k}} +\frac{\epsilon}{N}\Gamma_1 \sum_{j=1}^{N}({\bf \delta X_{\it j}}(t - \tau) - {\bf \delta X_{\it k}}(t))\nonumber\\ 
              &-&\epsilon\Gamma_2({\bf \delta X_{\it k}}(t) + {\bf \delta X^*}(t - \tau)),\;\;\;\;
              { {\it k}=1,2,...,{\it p}}   
              \label{msf1}     
\end{eqnarray}
\begin{eqnarray}
\delta {\dot{\bf X}_{\it l}}&=& {\bf J}f({\bf X_{\it s}})\delta {\bf X_{\it l}} +\frac{\epsilon}{N}\Gamma_1 \sum_{j=1}^{N}({\bf \delta X_{\it j}}(t - \tau) - {\bf \delta X_{\it l}}(t)),\nonumber \\
&&\;\;\;\;\;\;\;\;\;\;\;\;\;\;\;\;\;\;\;\;\;\;\;\;\;\;\;\;\;\;\;\; l=p+1, p+2,...,N
\label{msf2}
\end{eqnarray}
where ${\bf J}f$ denotes the {\it Jacobian} of $f$.

\par After clustered synchronization, reduce model from Eqs. (\ref{eq:sub_population_1}) and (\ref{eq:sub_population_2}) becomes,
\begin{eqnarray}
{\dot{\bf X}_{f}(t)}&=&f({\bf X_{f}(t)}) +\frac{\epsilon}{N}\Gamma_1 q({\bf X_s}(t - \tau) - {\bf X_{f}}(t))\nonumber\\ &+&\frac{\epsilon}{N}\Gamma_1 p({\bf X_f}(t - \tau) - {\bf X_{f}}(t)) \nonumber \\
              &-&\epsilon\Gamma_2({\bf  X_{\it f}}(t) + {\bf X_s}(t - \tau)),  
              \label{App_red1}     
\end{eqnarray}
and
\begin{eqnarray}
{\dot{\bf X}_{s}(t)}&=&f({\bf X_{s}(t)}) +\frac{\epsilon}{N}\Gamma_1 q({\bf X_{f}}(t - \tau) - {\bf X_{s}}(t)),\nonumber \\
&+&\frac{\epsilon}{N}\Gamma_1 p({\bf X_f}(t - \tau) - {\bf X_{s}}(t)), 
\label{App_red2}
\end{eqnarray}
respectively.
Eqs.\ (\ref{msf1}) and (\ref{msf2}) are the master stability equations for the cluster synchronization. Then we calculate two maximum Lyapunov exponents (MLEs) corresponding  to transverse direction of the cluster synchronization manifolds from master stability Eqs.\ (\ref{msf1}) and (\ref{msf2}) using Eqs.\ (\ref{App_red1}) and (\ref{App_red2}). Negativity of the MLEs by changing of coupling strength $\epsilon$ gives the necessary condition for cluster synchronization.


\begin{thebibliography}{40}
\bibitem{Zanette2005} D. H. Zanette, %Synchronization and frustration in oscillator networks with attractive and repulsive interactions, 
\textit{Europhys. Lett.} {\bf 72}, 190 (2005).

\bibitem{Strogatz-Hongprl} H. Hong and S. H. Strogatz, %Kuramoto Model of Coupled Oscillators with Positive and Negative Coupling Parameters: An Example of Conformist and Contrarian Oscillators, 
\textit{Phys. Rev. Lett.} {\bf 106}, 054102 (2011).

\bibitem{Chandrasekhar2018} K. Sathiyadevi, V. K. Chandrasekar, D. V. Senthilkumar, and M. Lakshmanan, %Distinct collective states due to trade-off between attractive and repulsive couplings, 
\textit{Phys. Rev. E} {\bf 97}, 032207 (2018).

\bibitem{Pikovsky} A. Yeldesbay, A. Pikovsky, and M. Rosenblum, % Chimeralike States in an Ensemble of Globally Coupled Oscillators,
 \textit{Phys. Rev. Lett.} {\bf 112}, 144103 (2014).

\bibitem{Mishra15} A. Mishra, C. R. Hens, M. Bose, P. K. Roy, and S. K. Dana, % Chimeralike states in a network of oscillators under attractive and repulsive global coupling,
 \textit{Phys. Rev. E} {\bf 92}, 062920 (2015).

\bibitem{Chen2009} Y. Chen, J. Xiao, W. Liu, Y. Yang, \textit{Phys. Rev. E} {\bf 80}, 046206 (2009).

\bibitem{Rabinovich2006} M.I. Rabinovich, P. Varona, A.I. Selverston, and H.D.Abarbanel, \textit{Rev. Mod. Phys.} {\bf 78}  1213 (2006).

\bibitem{A Turing:PTRSL_1952} A. Turing, %The chemical basis of morphogenesis,
 \textit{Philos. Trans. R. Soc. London} {\bf 237}, 37--72 (1952).

\bibitem{Nandan2014} M. Nandan, C. R. Hens, P. Pal, and S. K. Dana, %Transition from amplitude to oscillation death in a network of oscillators, 
\textit{Chaos} {\bf 24}, 043103 (2014). 

\bibitem{Zhao2018} N. Zhao, Z. Suna, and W. Xu, %Amplitude death induced by mixed attractive and repulsive coupling in the relay system, 
\textit{Eur. Phys. J. B} {\bf  91,}  20 (2018).

\bibitem{G Saxena:Phys Rep_2012} G. Saxena, A. Prasad and R. Ramaswamy, %
%{Amplitude death: The emergence of stationarity in coupled nonlinear systems}, 
{\textit Phys. Rep.} {\bf 521}, 205 (2012). 

\bibitem{A Koseska:Phys Rep_2013} A. Koseska, E. Volkov and J. Kurths, %{Oscillation quenching mechanisms: Amplitude vs. oscillation death},
 {\it Phys. Rep.} {\bf 531}, 173 (2013).
 
 

\bibitem{Matthews1990} P. C. Matthews and S. H. Strogatz, %{Phase diagram for the collective behavior of limit-cycle oscillators}, 
{\it Phys. Rev. Lett.} {\bf 65}, 1701 (1990)%; P. C. Matthews, R. E. Mirollo and S. H. Strogatz,
% {"Dynamics of a large system of coupled nonlinear oscillators"}, {\it Physica D} {\bf 52}, 293 (1991).

\bibitem{Aronsen1990} D.G. Aronson, G.B. Erementrout, N. Kopell, \textit{Physica D}, {\bf 41}, 403 (1990).


\bibitem{Konishi2003} K. Konishi, \textit{Phys. Rev. E} {\bf 68}, 13 (2003).

\bibitem{Karnatak2007} R. Karnatak, R. Ramaswamy, A. Prasad, \textit{Phys. Rev. E} {\bf 76}, 432 (2007).

\bibitem{Resmi2011} V. Resmi, G. Ambika, R.E. Amritkar, \textit{Phys. Rev. E} {\bf 84}, 046212 (2011).

\bibitem{Sen1998} D. V. R. Reddy, A. Sen, and G. L. Johnston, % Time Delay Induced Death in Coupled Limit Cycle Oscillators, 
\textit{Phys. Rev. Lett.} {\bf 80}, 5109, (1998).


%\bibitem{rayleigh} J. W. S. Rayleigh, The Theory of Sound (Dover, New York, 1945), Vol. 2; M. Abel, K. Ahnert, and S. Bergweiler, Phys. Rev. Lett. 103, 114301 (2009).



\bibitem{A Koseska:PRL_2013} A. Koseska, E. Volkov and J. Kurths, %{Transition from Amplitude to Oscillation Death via Turing Bifurcation}, 
{\it Phys. Rev. Lett.} {\bf 111}, 024103 (2013).

\bibitem{Hens2013} C.R. Hens, O.I. Olusola, P. Pal, S.K. Dana, % Oscillation death in diffusively coupled oscillators by local repulsive link, 
{\it Phys. Rev. E} {\bf 88} 034902 (2013);  C.R. Hens, P. Pal, S.K. Bhowmick, P.K. Roy, A. Sen, S.K. Dana, %Diverse routes of transition from amplitude to oscillation death in coupled oscillators under additional repulsive links,
 {\it Phys. Rev. E} {\bf 89} 032901 (2014).

\bibitem{Zou2013} W. Zou, D.V. Senthilkumar, A. Koseska, J. Kurths, %{Generalizing the transition from amplitude to oscillation death in coupled oscillators}, 
{\it Phys. Rev. E} {\bf 88}  050901 (2013); W. Zou, D. V. Senthilkumar, J. Duan, J. Kurths, %{Emergence of amplitude and oscillation death in identical coupled oscillators}, 
{\it Phys. Rev. E} {\bf 90}  032906 (2014), 


\bibitem{Banerjee2014_expt} T. Banerjee, D.Ghosh, %{Experimental observation of a transition from amplitude to oscillation death in coupled oscillators},
 {\it Phys. Rev. E}  {\bf 89}, 062902 (2014). 
 
\bibitem{Bera2016} B. K. Bera, C. R. Hens, S. K. Bhowmick, P. Pal, and D. Ghosh, % {Transition from homogeneous to inhomogeneous steady states in oscillators under cyclic coupling},
 \textit{Phys. Lett. A} {\bf 380}, 130-134, (2016).
\bibitem{pla2016} S. Majhi, B. K. Bera, S. K. Bhowmick, and D. Ghosh, \textit{Phys. Lett. A} {\bf 380}, 3617-3624, (2016).
\bibitem{Banerjee2014} T. Banerjee, D.Ghosh, %{Transition from amplitude to oscillation death under mean-field diffusive coupling},
{\it Phys. Rev. E} {\bf 89}, 052912(2014).
\bibitem{Suresh2016} D. V. Senthilkumar, K. Suresh, V. K. Chandrasekar, W. Zou, S. K. Dana, T. Kathamuthu, and J. Kurths, %Experimental demonstration of revival of oscillations from death in coupled nonlinear oscillators, 
\textit{Chaos} {\bf 89}, 043112 (2016).

\bibitem{Verma2017} U. K. Verma, A. Sharma, N. K. Kamal, J. Kurths, and M. D. Shrimali, %Explosive death induced by mean?field diffusion in identical oscillators, 
\textit{Sci. Rep.} {\bf 7}, 7936, (2017).


\bibitem{Bera2016_pla} B. K. Bera, C. R. Hens, and D. Ghosh, % Emergence of amplitude death scenario in a network of oscillators under repulsive delay interaction, 
\textit{Phys. Lett. A} {\bf 380}, 2366-2373, (2016).




\bibitem{Zou2013_prl} W. Zou, D. V. Senthilkumar, M. Zhan, and J. Kurths,
 {\it Phys. Rev. Lett.}  {\bf 111}, 014101 (2013). 
 \bibitem{Zou2015} W. Zou, D. V. Senthilkumar, R. Nagao, I. Z. Kiss, Y. Tang, A. Koseska, J. Duan, and J. Kurths, 
 {\it Nat. Comm.}  {\bf 6}, 7709 (2015). 
\bibitem{Nagao2016} R. Nagao, W. Zou, I. Z. Kiss and J. Kurths {\it Chaos.}  {\bf 26}, 094808 (2016). 
 
 \bibitem{Daido2004} H. Daido and K. Nakanishi 
 {\it Phys. Rev. Lett.}  {\bf 93}, 104101 (2004). 
 \bibitem{Kundu2018} S. Kundu, S. Majhi and D. Ghosh,  {\it Phys. Rev. E} {\bf 97}, 052313 (2018).
 \bibitem{Zou2009} W. Zou and M. Zhan, 
 {\it Phys. Rev. E}  {\bf 80}, 065204 (2009). 

\bibitem{auto} B. Ermentrout, % {``\it Simulating, Analyzing, and Animating Dynamical Systems: A Guide to Xppaut for Researchers and Students (Software, Environments, Tools),"} 
SIAM Press, Philadelphia, PA, 2002.
\bibitem{msf} L. M. Pecora, and T. L. Carroll, {\it Phys. Rev. Lett.}  {\bf 80}, 2109 (1998).


\end{thebibliography}
\end{document}